\documentclass[%
 reprint,
 amsmath,amssymb,
 aps,
]{revtex4-2}

\usepackage{graphicx}
\usepackage{dcolumn}
\usepackage{bm}

\newcommand{\mycomment}[1]{}
\usepackage{subfigure}

\begin{document}

\preprint{APS/123-QED}

\title{Ising model with non-reciprocal interactions}

\author{Agney K. Rajeev}
\affiliation{%
 School of Physical Sciences, National Institute of Science Education and Research, \\HBNI, Jatni, Bhubaneswar 752050, India
}

\author{A. V. Anil Kumar}%
 \email{anil@niser.ac.in}
\affiliation{%
 School of Physical Sciences, National Institute of Science Education and Research, \\HBNI, Jatni, Bhubaneswar 752050, India
}

\date{\today}

\begin{abstract}
Effective interactions that violate Newton's third law of action-reaction symmetry are common in systems where interactions are mediated by a non-equilibrium environment. Extensive Monte Carlo simulations are carried out on a two-dimensional Ising model, where the interactions are modified non-reciprocally. We demonstrate that the critical temperature decreases as the non-reciprocity increases and this decrease depends only on the magnitude of non-reciprocity. Further, travelling spin waves due to the local fluctuations in magnetisation are observed and these spin waves travel opposite to the non-reciprocity vector.

\end{abstract}

\maketitle


\section{Introduction}

In many of the complex systems, the interactions between the particles are mediated by the environment. For example, motile bacteria modify their activity based on the local concentration of autoinducer molecules, which is termed as quorum sensing\cite{whiteley}. This quorum sensing changes the effective interactions between the motile bacteria. Another example is active systems, such as bird flocks, with a visual perception. It has been established that the dynamics of each bird in a flock is affected by the dynamics of its seven nearest neighbours\cite{ballerini,pearce}. Such interactions are accounted for in recent models as visual perception\cite{barberis,lavergne,negi} and velocity alignment\cite{viscek,negi1}. These environment mediated interactions are generally non-reciprocal; i.e. the action-reaction symmetry is broken.  One major implication of non-reciprocal interactions is that these systems do not obey detailed balance and time reversal symmetry, thereby driving the system out of equilibrium. This results in the emergence of new features such as time-dependent states and oscillatory dynamics leading to self-propelling bands or dynamic micro patterns\cite{saha}.

These non-reciprocal interactions and the non-equilibrium dynamics generated by them play a significant role in the structure and dynamics of active particles\cite{Dinelli,stark,durve}, animal groups\cite{halutes}, social systems\cite{helbing,karamouzas}, complex plasmas\cite{chaudhuri,morfil}, biochemical systems\cite{kronzucker,bo} etc. Currently, these systems, motivated by the effect of non-reciprocal interactions, have become a focus of growing research interest. In general, 
the non-reciprocal interactions are non-Hamiltonian and most of the classical statistical mechanics formalism cannot be applied to systems with non-reciprocal interactions. Recently, there have been a large number of numerical and theoretical investigations to understand the effects of non-reciprocal interactions on the phase behaviour and dynamics of such non-equilibrium systems\cite{fruchart, duan, ivlev, banerjee, poncet,bartnick,mandal}. It has been shown that the interplay between non-reciprocal enhancement of fluctuations and many-body effects leads to time-dependent phases which can be viewed as dynamical restorations of spontaneously broken continuous symmetry\cite{fruchart}. The interplay between self-propulsion and non-reciprocity  induces  a chaotic chasing band phase\cite{duan}. Another interesting observation about these time-dependent phases is that these systems can give rise to stable travelling waves\cite{saha,you}. Some of the predictions from these investigations were verified experimentally on two-dimensional binary complex plasmas\cite{ivlev}, catalytically active colloids\cite{soto} as well as in molecular systems\cite{mandal1}. 

In this work, we introduce an Ising model with non-reciprocal interactions and carry out detailed Monte Carlo investigations on the model. Motivated by recent studies on the binary mixtures of particles, we introduce non-reciprocal interactions by changing the interaction strength between the nearest neighbours non-reciprocally. We observe that the critical temperature decreases parabolically as the non-reciprocity increases. We have also observed travelling waves whose direction is determined by the direction in which non-reciprocity is applied. The paper is organised as follows: Section II introduces the non-reciprocal Ising model and simulation details. Numerical results are presented in section III together with the discussions. Finally, we summarize this paper in section IV.

\section{Model \& Methods}

We have considered the simple two-dimensional square lattice Ising model with nearest neighbour interactions on \textbf{L} $\times$ \textbf{L} lattices with periodic boundary conditions. Each lattice point $i \in [1,L]^{2} \cap \mathbb{Z} ^{2}$ is attributed with a spin variable $\sigma_{i} \in \{-1,1\}$.
In the absence of any external field, the interaction Hamiltonian for the Ising model is given by 

\begin{equation}
    {\cal H} = - J \sum_{\langle i,j \rangle}\sigma_{i} \sigma_{j}
\end{equation}

\noindent where $\langle i, j \rangle$ denotes distinct nearest neighbour pairs and $J$ is the interaction strength. To introduce non-reciprocal interactions, we have modified the Hamiltonian as follows. The Hamiltonian of a spin variable $\sigma_{i}$ is given by the equation

\begin{equation}
    {\cal H}_{i} = - \sigma_{i} \sum_{j} J_{ij} \sigma_{j}
\end{equation}

\noindent where $j \in [1,L]^{2} \cap \mathbb{Z} ^{2}$ and $J_{ij}$ is called the \textit{interaction parameter}. The traditional Ising model considers only the interactions of a spin with its nearest neighbours and has an interaction parameter of the form

\begin{equation}
    J_{ij} = J[\delta_{i-\hat{e_{x}}, j} + \delta_{i+\hat{e_{x}}, j} + \delta_{i-\hat{e_{y}}, j} + \delta_{i+\hat{e_{y}}, j}]
\end{equation}

\noindent where $J$ is a constant implying equal interactions among all nearest neighbours. Also any lattice point $i \equiv (i_{x}, i_{y}) \equiv i_{x}\hat{e_{x}} + i_{y}\hat{e_{y}}$ where $\hat{e_{x}}$ and $\hat{e_{y}}$ are the unit vectors in x and y direction respectively.
It can be noted that $J_{ij} = J_{ji}$, implying a reciprocal interaction between the spins $\sigma_{i}$ and $\sigma_{j}$. 

\par In many of the investigations, the non-reciprocal interactions are introduced by choosing the interaction strength parameter to be different for particle pairs. For example, in a binary mixture of particles, the non-reciprocity in the force between the particles is given by\cite{ivlev}

\begin{equation}
  \boldsymbol{F}_{ij}=-\frac{\partial \phi(r_{ij})}{\partial \boldsymbol{r}_j}\begin{cases}
    1-\Delta, & \text{for $ij \in AB$}.\\
    1+\Delta, & \text{for $ij \in BA$}.\\
    1, & \text{for $ij \in AA$ or $BB$}.
  \end{cases}
\end{equation}

Taking a cue from these studies, we incorporate non-reciprocal interactions into our model by modifying the interaction parameter as follows:

\begin{eqnarray}
    J_{ij} = &&J[(1 + \Delta_{x})\delta_{i-\hat{e_{x}}, j} + (1 - \Delta_{x})\delta_{i+\hat{e_{x}}, j} 
    \nonumber\\ && + \;(1 + \Delta_{y})\delta_{i-\hat{e_{y}}, j} + (1 - \Delta_{y})\delta_{i+\hat{e_{y}}, j}]
\end{eqnarray}

\noindent Here, $\Delta_{x}, \Delta_{y} \in [-1,1]$ and can be written as $\Delta_{x}\hat{e_{x}} + \Delta_{y}\hat{e_{y}} \equiv (\Delta_{x}, \Delta_{y}) \equiv \boldsymbol{\Delta}$. We also take the constant $J = 1$. This makes $J_{ij} \neq J_{ji}$ and hence the interactions become non-reciprocal.  Please note that our non-reciprocal Ising model is different from that outlined in \cite{seara}, 
where the strength of the interaction parameter is kept constant, but the spins with which the given spin is interacting can be non-nearest neighbours. In our model, we kept the interactions with nearest neighbours, but the interaction parameter is varied. Also, our model is different from the active Ising model, since our model does not include any self-propulsion.

We now place the two-dimensional square lattice of Ising spins with non-reciprocal interactions in contact with a temperature bath at temperature $T$. The system is evolved using the Metropolis algorithm by flipping random spins in the lattice and accepting/rejecting the spin-flip with a probability determined by the difference in energy of the flipped spin. If $[\sigma]$ is a spin configuration of the lattice and $[\sigma]_{i}$ is the spin configuration after flipping the $i^{th}$ spin $\sigma_{i}$ of $[\sigma]$ to $-\sigma_{i}$, then the probability of accepting the spin-flip (and hence, $[\sigma_{i}]$) is given by
\begin{equation}
    P ([\sigma] \to [\sigma]_{i}) = min\{1,e^{-\frac{\Delta E_{i}}{T}}\}
\end{equation}

\noindent where $\Delta E_{i} = {\cal H}_{i}([\sigma]_{i}) - {\cal H}_{i}([\sigma])$. Please note that the change in energy of the individual spin $\Delta E_i$ is not equal to the change in the total system energy due to the spin-flip.  This happens due to the non-reciprocal nature of the interactions. Thus detailed balance is not satisfied and our model is non-equilibrium. Similar models were used to study dynamics in directed networks\cite{sanchez,lipowski}.

We have carried out extensive Monte Carlo simulations of this model for different values of $\boldsymbol{\Delta}$.  The system is allowed to evolve using the Monte Carlo acceptance criteria outlined above until, after some initial transient states, a stationary steady state is achieved. The various thermodynamic quantities are calculated and averaged over. These quantities are monitored to determine the possible phase transitions and Transition temperatures.

\section{Results}

\subsection{Transition temperature versus non-reciprocity}

As a first step in our analysis, we study the model with non-reciprocal interactions only along the x-axis ($\boldsymbol{\Delta} = \Delta \hat{e_{x}}$). We have calculated the following thermodynamic parameters to determine the transition temperature in this model. $E$ and $M$ are the canonical thermodynamical averages of energy and magnetisation per spin at temperature $T$, as given by 

\begin{equation}\label{5}
    E = \frac{\langle H \rangle}{N}
\end{equation}
\begin{equation}\label{6}
    M = \frac{\langle S \rangle}{N}
\end{equation}
\noindent where $H = \frac{1}{2}\sum_{i}{\cal H}_{i}$ and $S = \sum_{i}\sigma_{i}$. The specific heat capacity, $C$, and magnetic susceptibility $\chi$ are thus given by

\begin{equation}\label{7}
    C = \frac{\langle H^{2} \rangle - \langle H \rangle^{2}}{T^{2}}
\end{equation}
\begin{equation}\label{8}
    \chi = \frac{\langle S^{2} \rangle - \langle S \rangle^{2}}{T}
\end{equation}

\begin{figure}[h]
    \centering
    \hspace*{-0.9cm} 
    \includegraphics[scale=0.1]{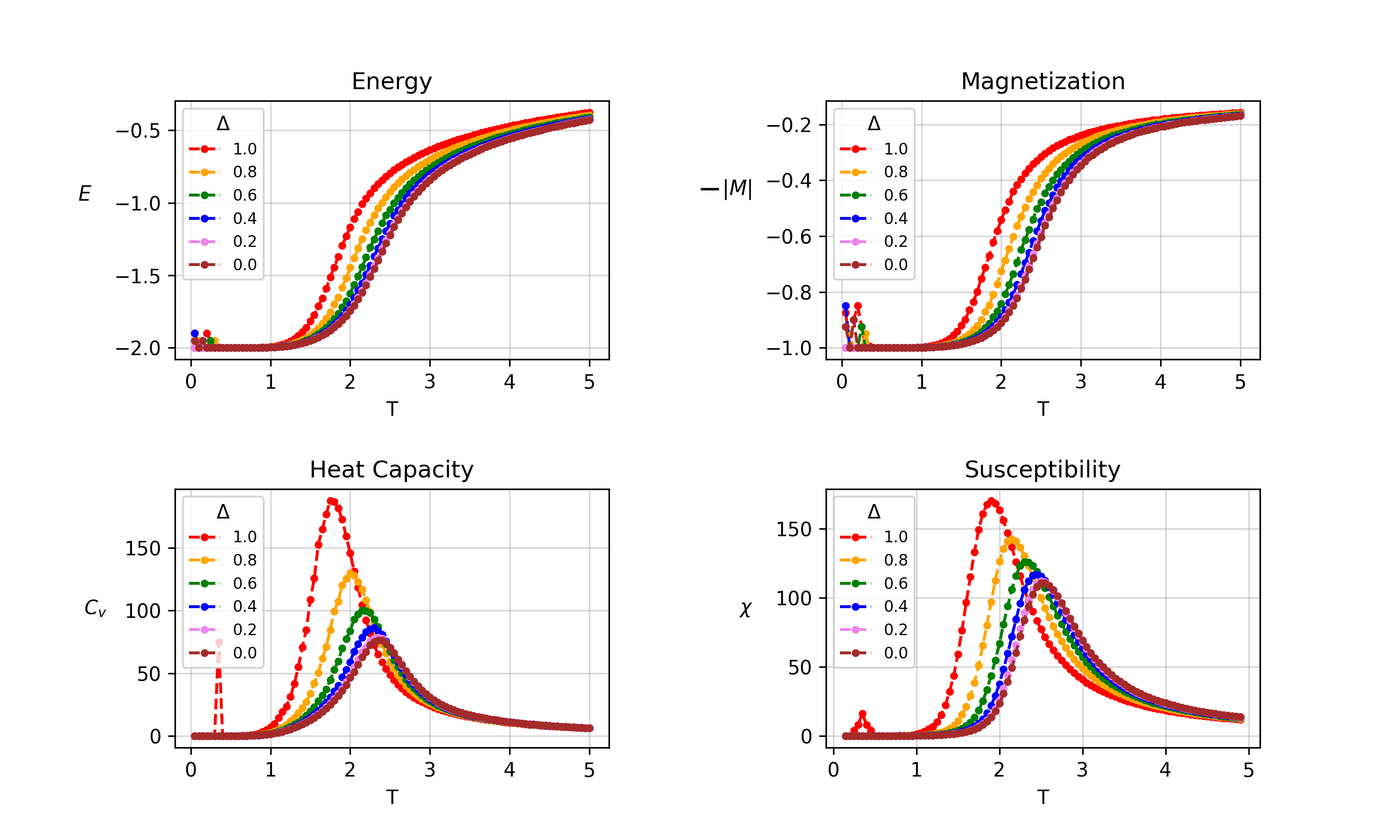}
    \caption{Plots of the evolution of average values of energy, magnetisation, heat capacity and susceptibility with temperature for various values of $\Delta$ with $L=8$ and $10^{7}$ spin flips per temperature}
    \label{fig:polished}
\end{figure}

\begin{figure}[h]
    \centering
    \includegraphics[scale=0.5]{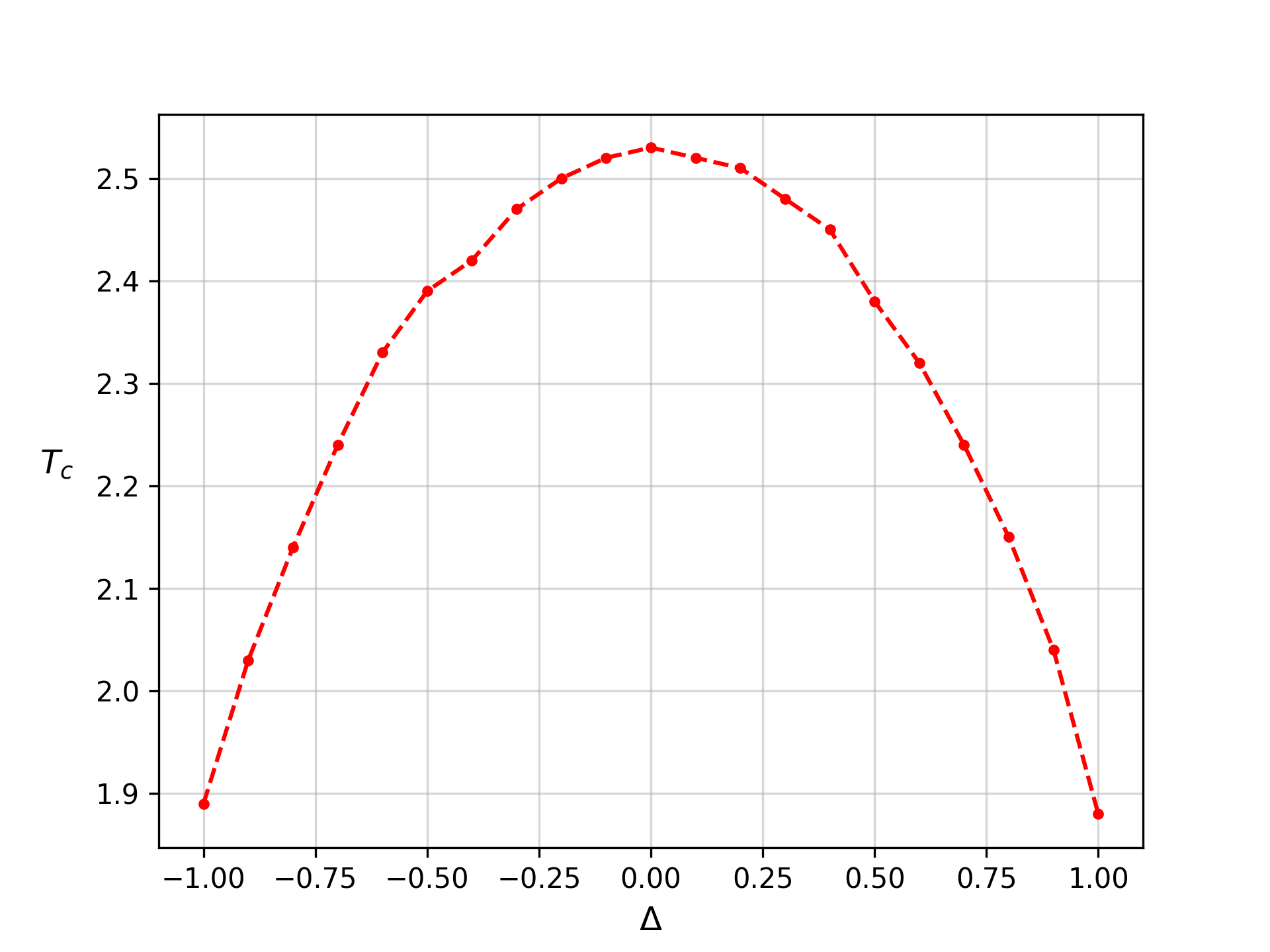}
    \caption{Plot of $T_{c}$ vs $\Delta$ obtained from fig. \ref{fig:polished}}
    \label{fig:two}
\end{figure}

The average values of these thermodynamic quantities at different values of $\boldsymbol{\Delta}$ are plotted in Figure \ref{fig:polished} for an\\ 8 $\times$ 8 lattice. From the figure, it is evident that this model shows a continuous phase transition from disordered to ordered phase akin to the traditional Ising model and the critical temperature $T_c$ can be obtained. For $\boldsymbol{\Delta} $ = 0, the transition temperature is close to 2.55, which is larger than the theoretical value of 2.269. We denote this as $T_{c0}$. However, when we do finite size scaling, which will be described later, the transition temperature of the macroscopic system is reduced to 2.286, which is very close to the theoretical value.  As $\boldsymbol{\Delta} $ increases,  the peaks in $C$ and $\chi$ shift towards lower values of temperature indicating a lower critical temperature. Thus,  there is a notable shift in the critical temperature($T_{c}$) of the transition towards lower values.  This shift in the critical temperature becomes larger as we increase $\boldsymbol{\Delta} $.  The critical temperature for different $\boldsymbol{\Delta} $ is plotted in Figure \ref{fig:two}. $T_c$  is maximum at $\boldsymbol{\Delta}$ = 0 and reduces symmetrically with increasing modulus of $\boldsymbol{\Delta}$.  This reduction in $T_c$ as $\Delta = |\boldsymbol{\Delta}|$ increases, is reasonable, as non-reciprocity in essence introduces more disorder into the system thereby delaying the transition to the ordered state to a lower temperature. This is in contrast with the non-reciprocal model introduced by Seara {\it et al.}\cite{seara}, where they found that the transition temperature increases as the non-reciprocity increases.  Since we are applying $\boldsymbol{\Delta}$  only along the $x$-axis, our model has a reflection symmetry perpendicular to the $x$-axis.  Thus the  $T_c$ versus $\boldsymbol{\Delta}$  plot becomes symmetric.  

\begin{figure}[h]
\centering
\begin{subfigure}
    \centering
    \includegraphics[scale=0.25]{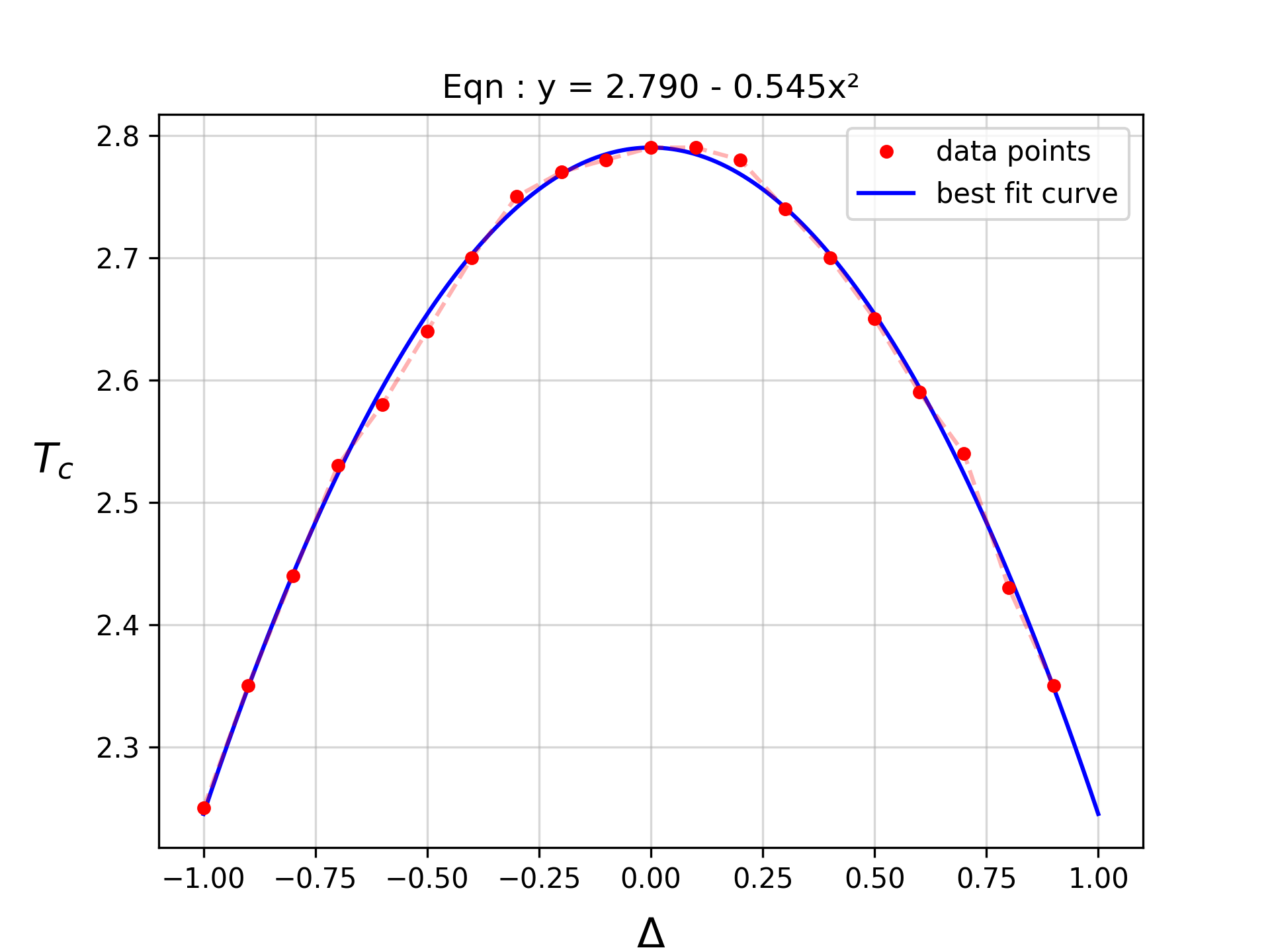}
    \label{fig:enter-label}
\end{subfigure}
\begin{subfigure}
    \centering
    \includegraphics[scale=0.25]{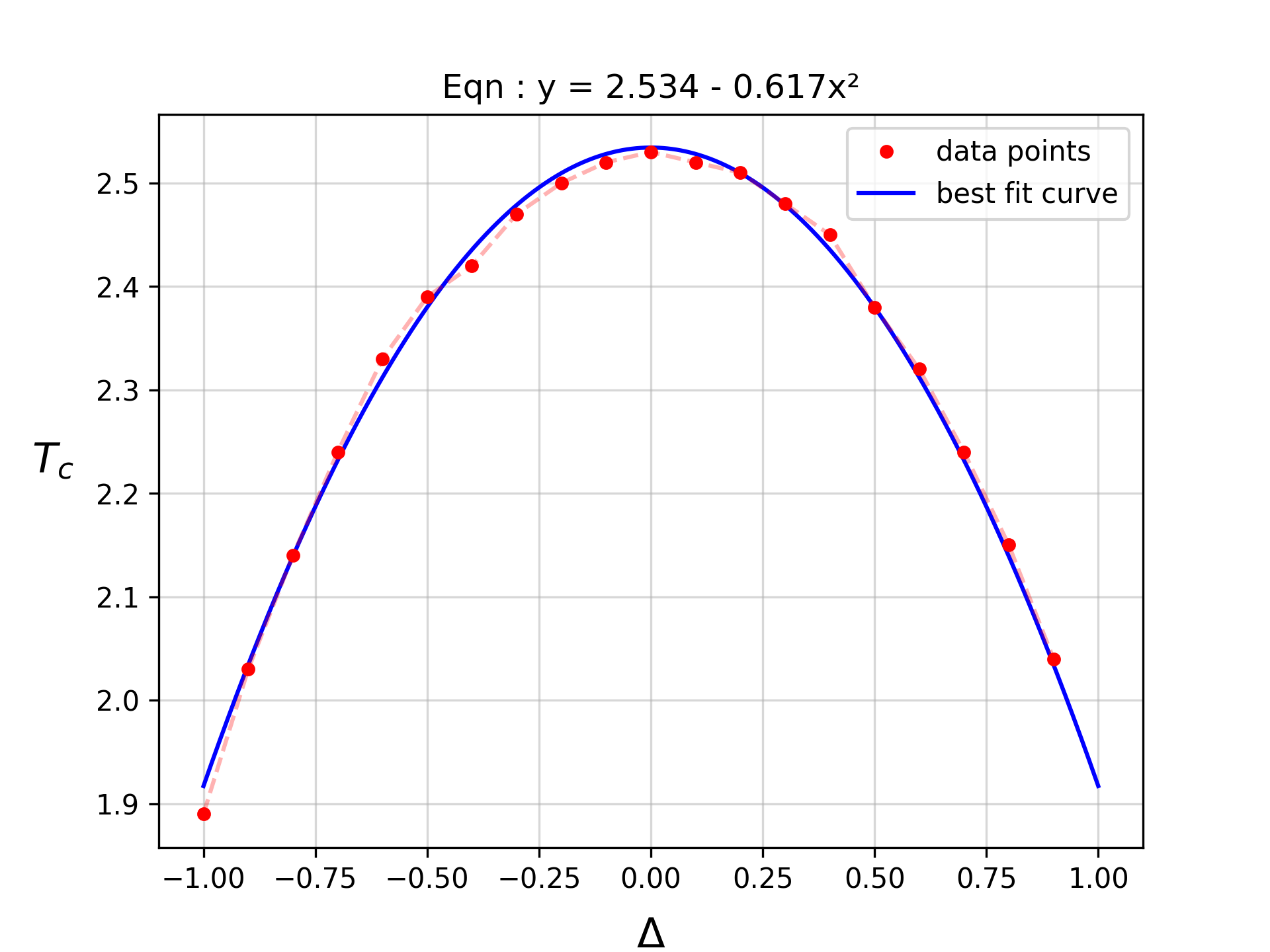}
\end{subfigure}
\begin{subfigure}
    \centering
    \includegraphics[scale=0.25]{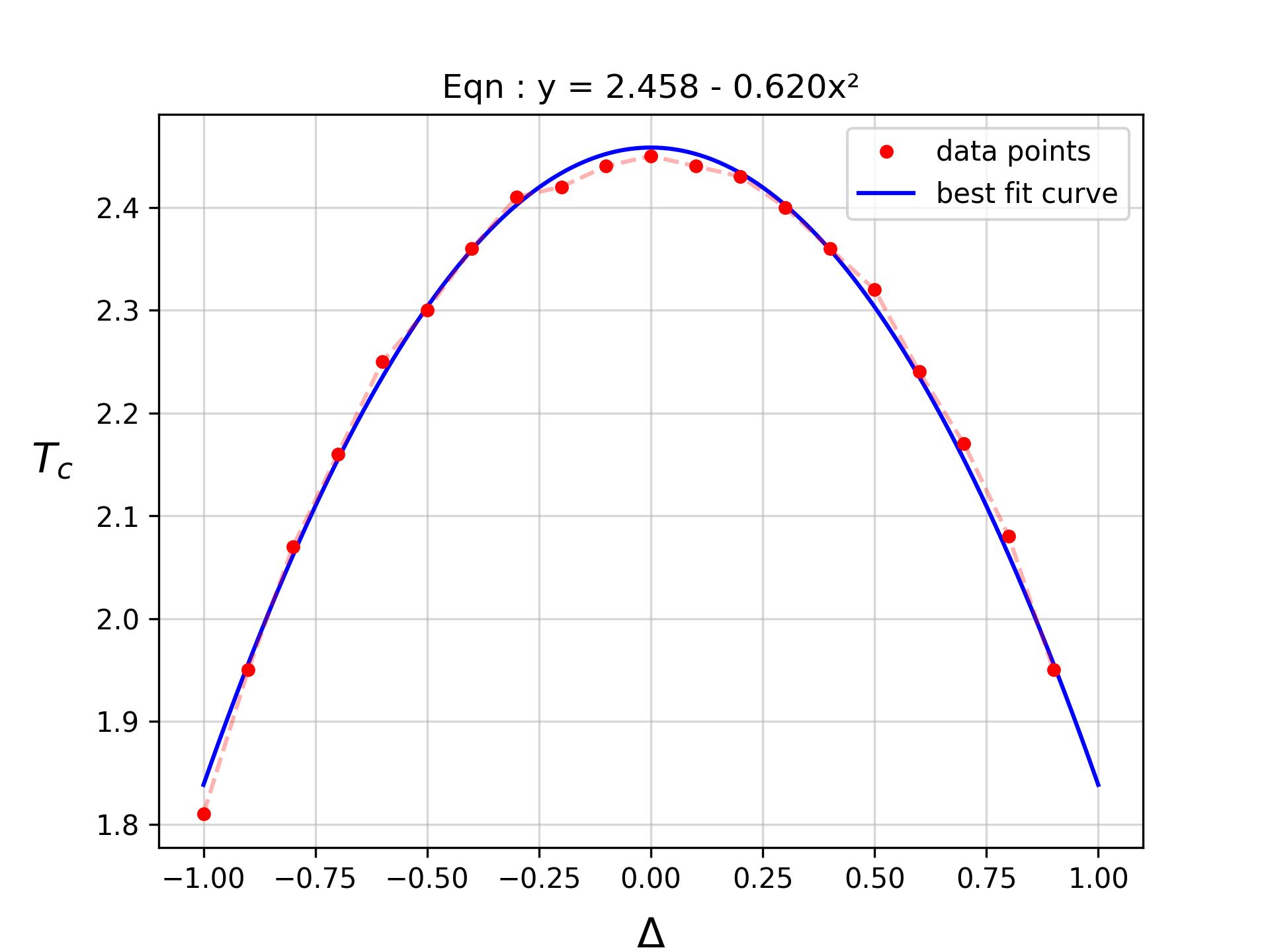}
\end{subfigure}
\begin{subfigure}
    \centering
    \includegraphics[scale=0.25]{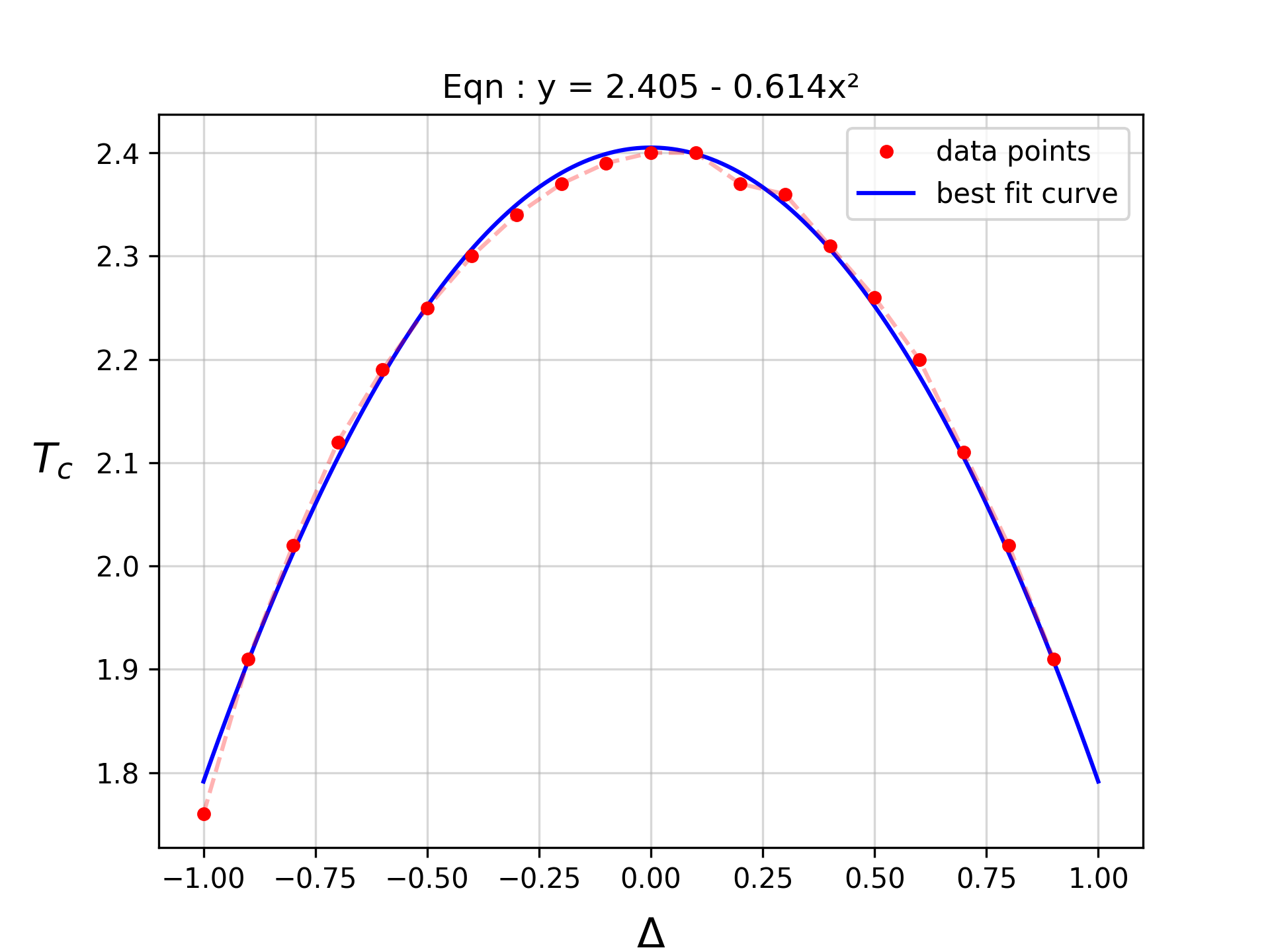}
    \label{fig:enter-label}
\end{subfigure}

\caption{$T_{c}$ vs $\Delta$ plot for $L=4,8,12,16$ respectively along with their best fit curve for eq \ref{eq:11}. The equation of the curve is noted on top}
\label{fig:quad_fit}
\end{figure}

 We have carried out Monte Carlo simulations for different lattice sizes to determine  that  these observations are valid irrespective of the system sizes. For all the lattice sizes in which Monte Carlo simulations have been carried out, we have observed that the plot of transition temperature versus the non-reciprocity parameter is symmetric about $\Delta$ = 0 and the value of $T_c$ reduces as the magnitude of $\Delta$ increases. These are plotted in Figure \ref{fig:quad_fit} for a few representative lattice sizes. We have tried to characterise this change in critical temperature with respect to $\Delta$ by applying a polynomial fit to these plots. The best fit yields a parabolic equation of the form 

\begin{equation}
    T_{c} = T_{c0} - k\Delta^{2}
    \label{eq:11}
\end{equation}
where $T_{c0}$ is the critical temperature at $\Delta=0$ and k is a proportionality factor. The parabolic form is found to be true for all lattice sizes. Since the energy parameter changes by $J(1 \pm \Delta)$, the first order corrections in energy will cancel each other and the dominant term will depend on $\Delta^2$ and this in turn will make the dependence of critical temperature on $\Delta^2$. So it is understandable that the change in $T_c$ will be of parabolic in nature.

\begin{figure}[!ht]
	\centering
	\hspace*{-0.45cm} 
	\includegraphics[scale=0.1]{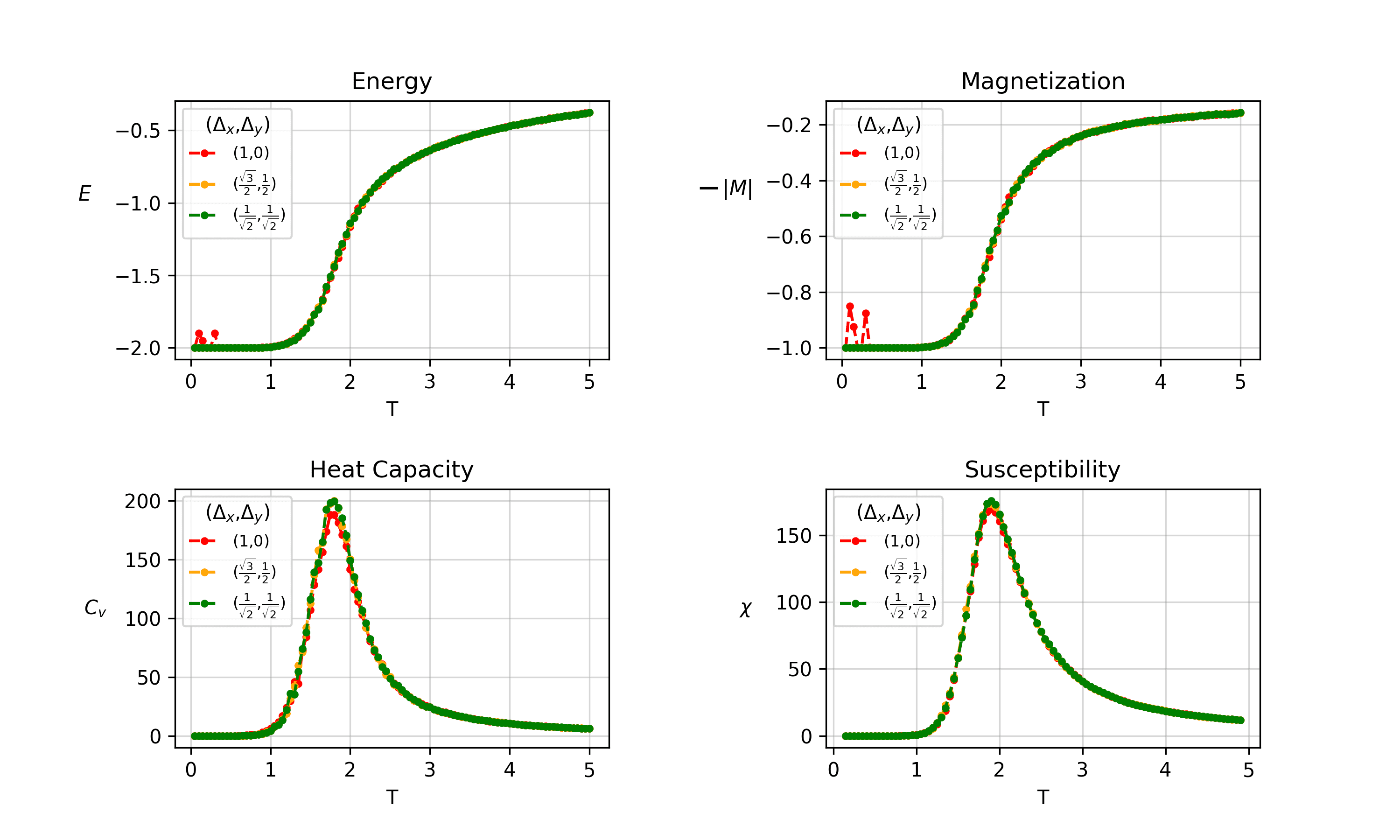}
	\caption{Plots of evolution of observables with temperature for various $\boldsymbol{\Delta} = (1,0),(\frac{\sqrt{3}}{2},\frac{1}{2}),(\frac{1}{\sqrt{2}},\frac{1}{\sqrt{2}})$ with $L=8$ and $10^{7}$ spin flips per temperature}
	\label{fig:six}
\end{figure}

The parabolic form of equation 11 suggests that change in $T_c$ depends only on the modulus of $\boldsymbol{\Delta}$ rather than the vector itself. In order to verify this, we have carried out Monte Carlo simulations by changing $\boldsymbol{\Delta}$ to $\Delta_{x} \hat{e_{x}} + \Delta_{y} \hat{e_{y}}$ to incorporate non-reciprocal interactions along both axes.  We have chosen 3 different $\boldsymbol{\Delta}$ vectors having unit magnitude , namely $\boldsymbol{\Delta} = (1,0),(\frac{\sqrt{3}}{2},\frac{1}{2}),(\frac{1}{\sqrt{2}},\frac{1}{\sqrt{2}})$ in these simulations. The average values of thermodynamic observables with change in temperature are plotted in Figure \ref{fig:six}. As evident from the figure, the values of physical observables for these three $\boldsymbol{\Delta}$ vectors coincide at all temperatures and the transition temperature remains the same for all the three $\boldsymbol{\Delta}$ vectors. As suggested above, this is consistent with equation 11, which says the change in critical temperature depends only on the absolute magnitude of $\boldsymbol{\Delta}$, and not its direction. However, as we discuss later on, the direction of the non-reciprocity vector indeed affects the direction of travelling states.
 
 \subsection{Finite size scaling}
 
 Finite size scaling theory is used to extrapolate the values of thermodynamic parameters obtained for finite systems.
 The transition temperature $T_c$ defined for infinite systems can be related to the transition temperature $T_c(L)$ defined for finite systems as\cite{fisher,privman} 
 \begin{equation}
  T_c - T_c(L) \sim L^{-1/\nu}
 \end{equation}

\begin{figure}[h]
\begin{subfigure}
    \centering
    \includegraphics[scale=0.45]{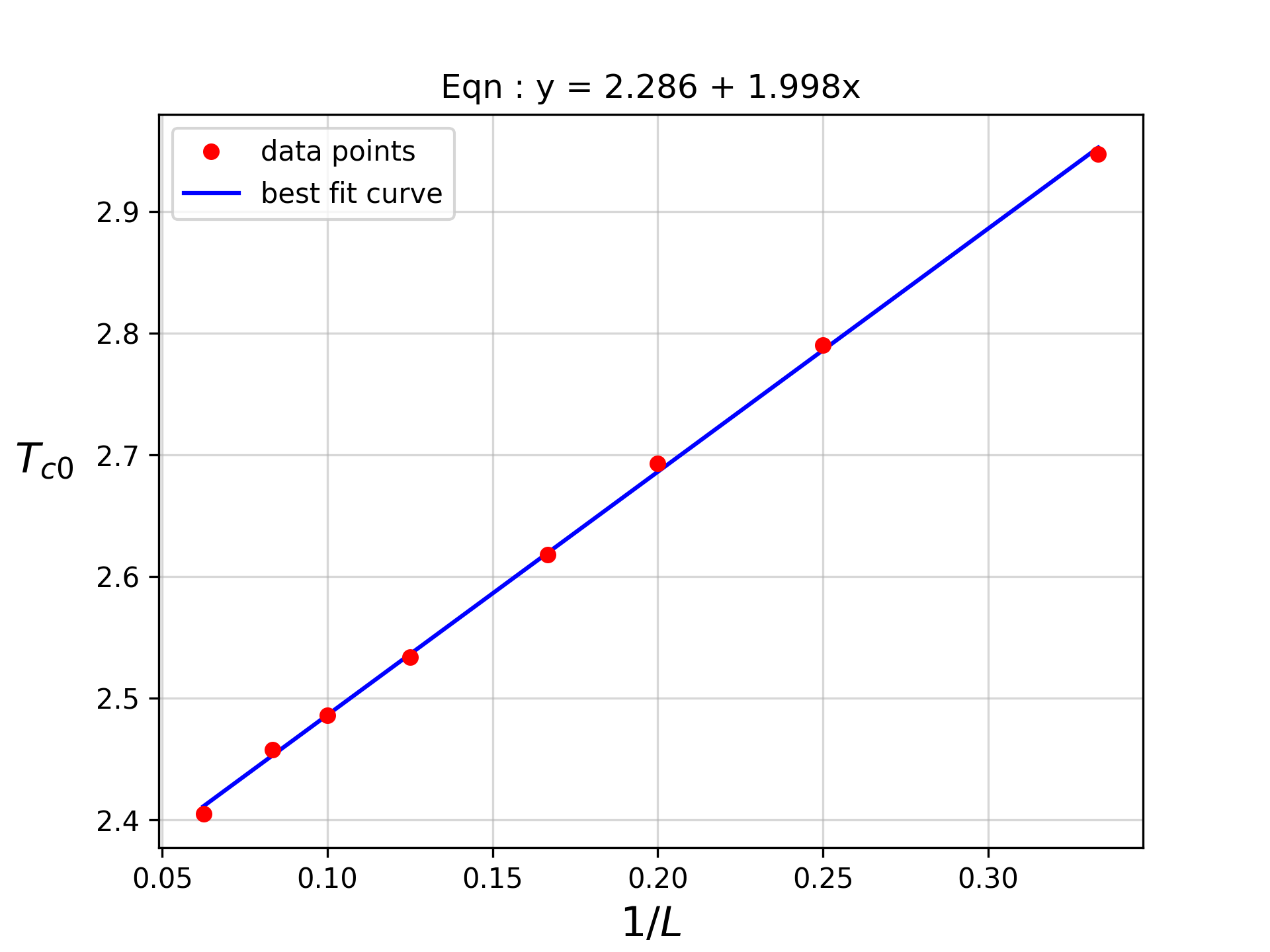}
    \label{fig:enter-label}
\end{subfigure}
\begin{subfigure}
    \centering
    \includegraphics[scale=0.45]{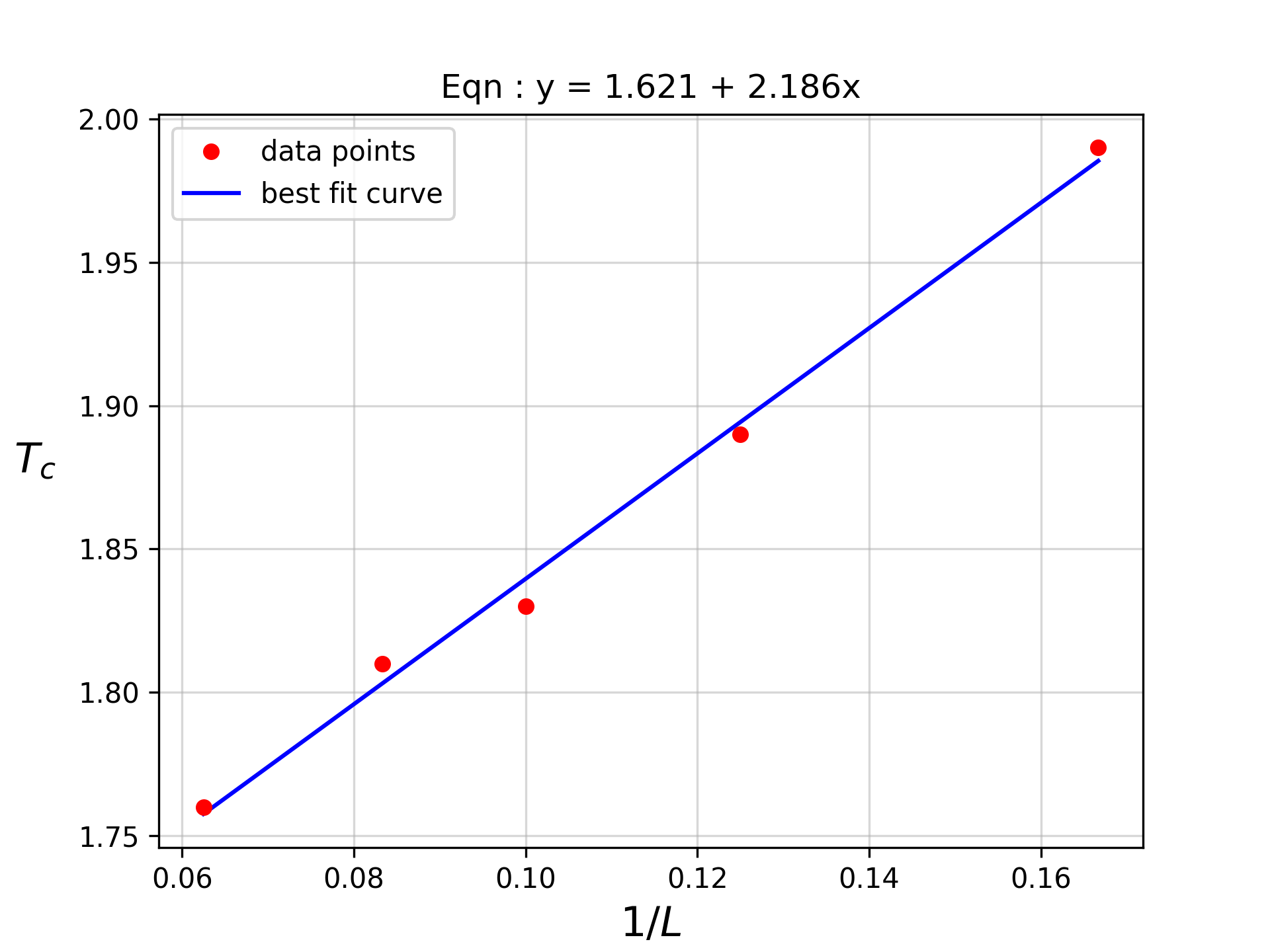}
    \label{fig:enter-label}
\end{subfigure}
    \caption{$T_{c}$ vs $1/L$ plot with best fit line for $\Delta$ = 0 and 1}
\label{fig:four}
\end{figure}

 We have carried out finite size scaling for $T_c$ to determine the bulk transition temperature. At $\Delta$ = 0, our model becomes the traditional Ising model. In Figure \ref{fig:four}, we have plotted the critical temperature $T_{c0}$ against 1/$L$. The points can be fitted to a linear relationship with a y-intercept which yields $T_{c}|_{L \to \infty}$ equal to  2.286 which is close to the analytical value of the critical temperature of the traditional ising model, 2.269. Also, this linear relationship provides us the correlation critical exponent, $\nu$ to be 1, which again agrees with the theoretical value for the two-dimensional Ising model.

\begin{figure}[!ht]
	\centering
	\hspace*{-0.6 cm}
	\includegraphics[scale=0.6]{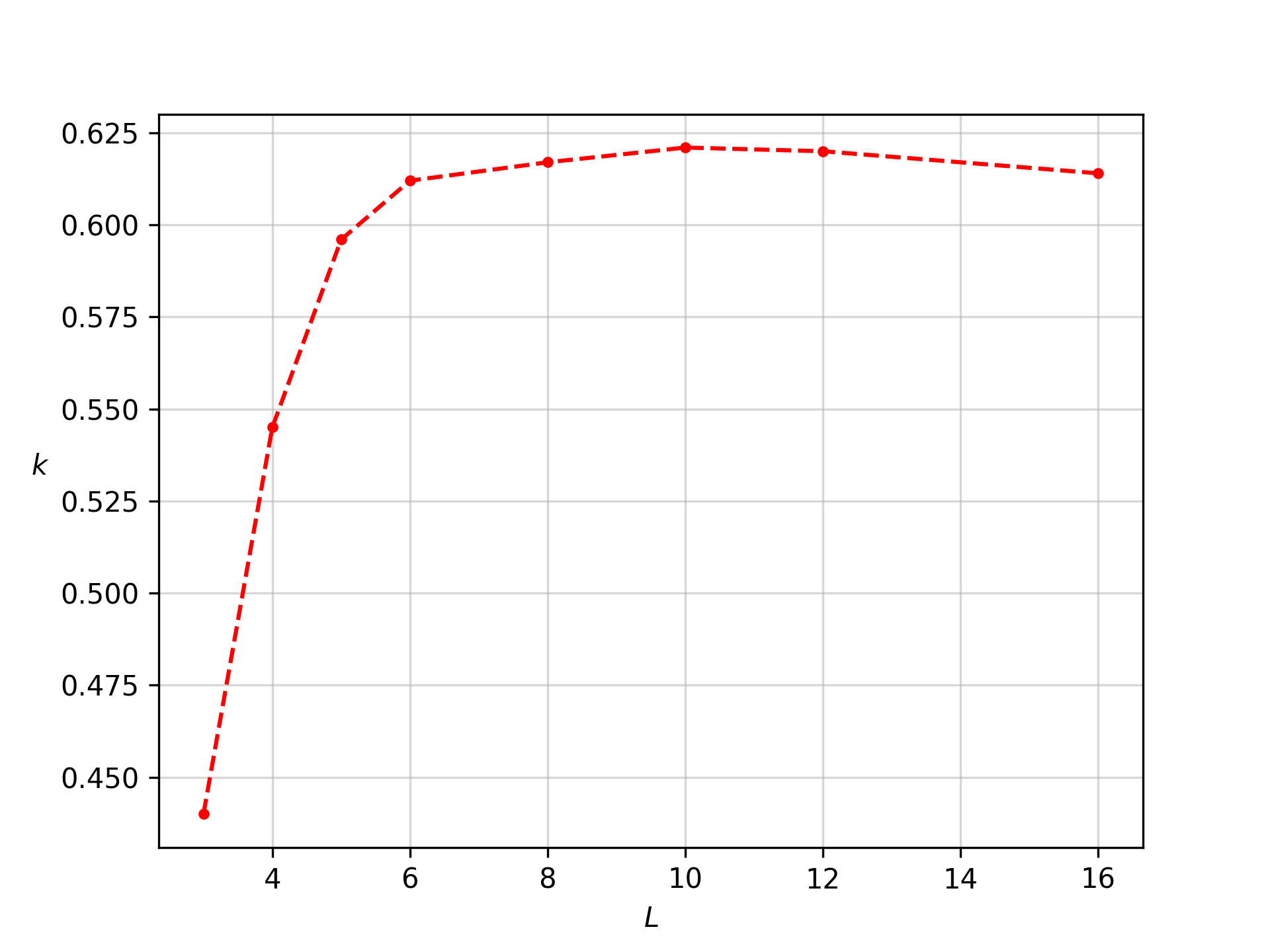}
	\caption{$k$ vs $L$ plot}
	\label{fig:five}
\end{figure}

We have applied the same finite size scaling to non-zero values of $\Delta$ and found that it works satisfactorily for them as well. As a typical example, we have plotted the $T_c(L)$ versus 1/$L$ for $\Delta$ = 1 (Note that only values of $L$ after saturation of $k$ in Fig. \ref{fig:five} were chosen). The data points can again be fitted with a straight line and the $T_c$ for the infinite lattice is given by the $y$-intercept as 1.621 which is very close to the value of $ T_{c0} - k $ = 1.666 and hence, is in agreement with eq. \ref{eq:11}. This suggests that the correlation exponent $\nu$ does not change if the interactions are non-reciprocal. 

 As stated above, the change in $T_c$ with $\Delta$ can be fitted with a parabolic form, given in equation 11. This form 
 remains unaltered for the lattice sizes we have simulated. we have checked the size dependence of the proportionality factor, $k$ of the parabolic equation. This is plotted in Figure \ref{fig:five}. As evident from the figure, the $k$ value increases initially as the lattice size increases, but saturates to a value $\sim$ 0.62 rapidly. 

\begin{figure}[!ht]
	\begin{subfigure}
		\centering
		\hspace*{-0.5cm} 
		\includegraphics[scale=0.21]{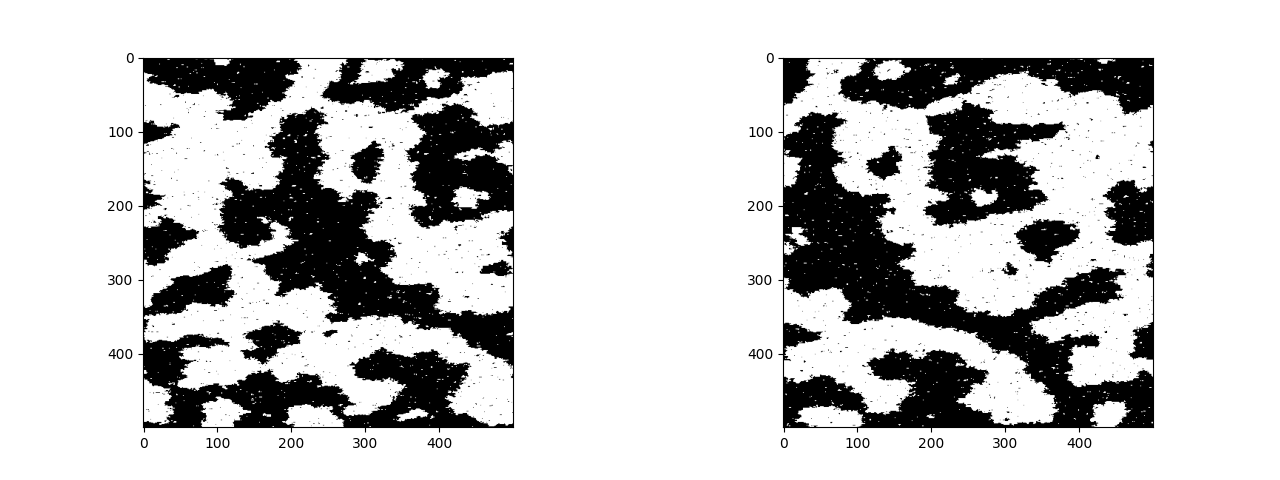}
	\end{subfigure}
	\begin{subfigure}
		\centering
		\hspace*{-0.5cm} 
		\includegraphics[scale=0.21]{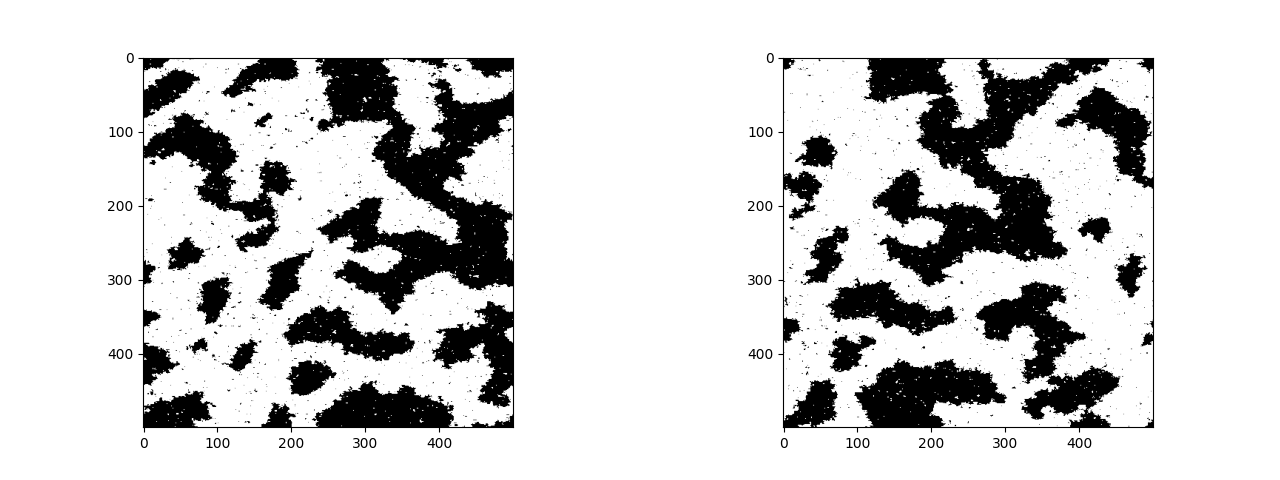}
	\end{subfigure}
	\begin{subfigure}
		\centering
		\hspace*{-0.5cm} 
		\includegraphics[scale=0.21]{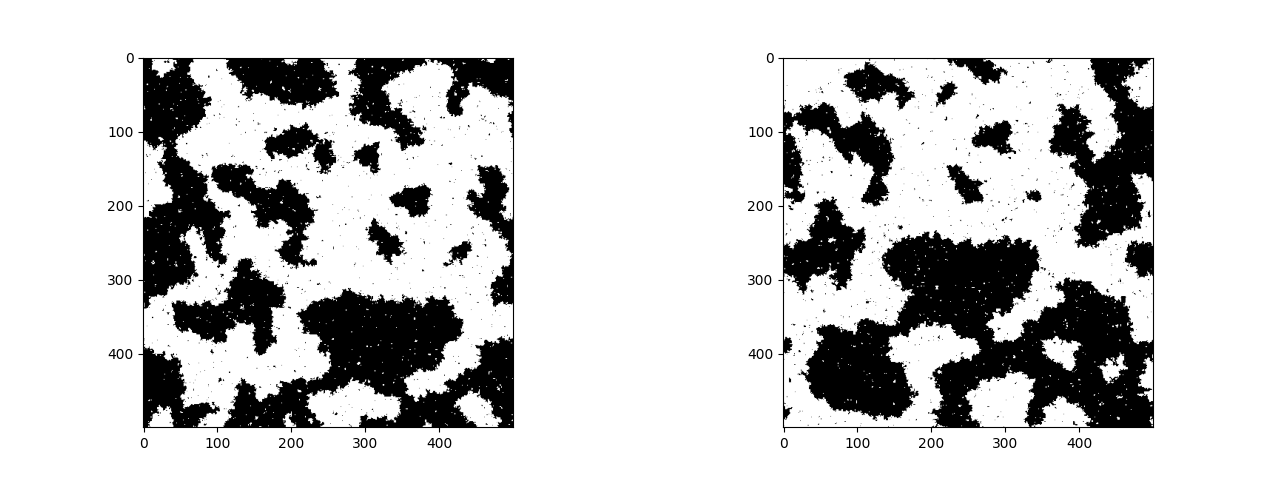}
	\end{subfigure}
	\caption{Comparison of Lattices ($L$=$500$, total no. of spin flips = $2.5 \times 10^{8}$) at the $1.995 \times 10^{8}$th and $2.495 \times 10^{8}$th spin flip for $\boldsymbol{\Delta} = (1,0),(\frac{\sqrt{3}}{2},\frac{1}{2}),(\frac{1}{\sqrt{2}},\frac{1}{\sqrt{2}})$ respectively. The shift in the clusters is indicative of the direction of the travelling wave (Note that the x-axis runs from left to right and the y-axis runs from top to bottom).}
	\label{fig:eight}
\end{figure}

\subsection{Travelling waves}

 One of the striking observations of systems with non-reciprocal interactions is the emergence of travelling patterns or waves in the steady state\cite{you,mandal,saha}. For $\boldsymbol{\Delta} >$ 0, we observe that the spin domains are travelling (Supplementary movies 1,2 and 3). In active particle systems with non-reciprocal interactions, these travelling waves are observed as the movement of clustered or flocked particles collectively. However, in our model system, the spins are attached to the lattice points. So the travelling spin waves we observe are due to the fluctuations in local magnetization. This is visible in Figure \ref{fig:eight}, where we compared the temporal evolution of our model at different spin flips for $\Delta = 1$ at two different Monte Carlo sweeps. This has been done on a 500 $\times$ 500 lattice. As earlier, we have chosen 3 different vectors $\boldsymbol{\Delta} = (1,0),(\frac{\sqrt{3}}{2},\frac{1}{2}),(\frac{1}{\sqrt{2}},\frac{1}{\sqrt{2}})$. In all the 3 systems, we observe the travelling waves. Moreover, as evident from the figure, the direction of the travelling waves is opposite to that of the $\boldsymbol{\Delta}$ vector (Note that the x-axis runs from left to right and the y-axis runs from top to bottom in Figure \ref{fig:eight}).

\section{Conclusions}

 We have introduced a non-reciprocal Ising model by varying the interaction strength non-reciprocally, inspired by earlier studies of particle systems where the interaction strength parameter is varied non-reciprocally. Extensive Monte Carlo simulations on this non-reciprocal Ising model are carried out to determine the effect of non-reciprocity on the paramagnetic to ferromagnetic transition. It has been observed that the transition temperature decreases monotonically as the strength of non-reciprocity increases. The dependence of $T_c$ on $\Delta$ is parabolic on the absolute value of the non-reciprocity parameter and does not depend on the direction in which non-reciprocity is introduced. We have also observed travelling spin waves, whose direction is opposite to the non-reciprocity parameter. From finite size scaling analysis, we observed that the correlation length exponent remains the same as the one in the non-reciprocal Ising model and does not have any dependence on the non-reciprocity parameter. It will be interesting to investigate how other critical exponents behave in the presence of non-reciprocal interactions. Our results indicate that rich and complex behaviour of non-reciprocal interactions can arise in simple lattice models and can be used as a model system to study the emergent phenomena in systems with non-reciprocal interactions. 

\begin{acknowledgments}
The authors acknowledge financial support from the
Department of Atomic Energy, India through the plan
Project (RIN4001-SPS).
\end{acknowledgments}

\end{document}